# *A long-term frequency-stabilized erbium-fiber-laser-based optical frequency comb with an intra-cavity electro-optic modulator

Zhang Yan-Yan(张颜艳)[a], Yan Lu-Lu(闫露露)[a], Zhao Wen-Yu(赵文宇)[a,b], Meng Sen(孟森)[a,c], Fan Song-Tao(樊松涛)[a,b], Zhang Long(张龙)[a], Guo Wen-Ge(郭文阁)[c], Zhang Shou-Gang(张首刚)[a] and Jiang Hai-Feng(姜海峰)[a]†

a) Key laboratory of Time and Frequency Standards, National Time Service Center, Xi'an 710600, China
b) Graduate University of Chinese Academy of Sciences, Beijing, 100039, China
c) School of Science, Xi'an Shiyou University, Xi'an, 710065, China

We demonstrate an optical frequency comb based on an erbium-doped-fiber femtosecond laser with the nonlinear polarization evolution scheme. The repetition rate of the laser is about 209 MHz. By controlling an intra-cavity electro-optic modulator and a piezo-transducer, the repetition rate can be stabilized with high servo bandwidth in a range of 3 kHz, enabling long-term repetition rate phase-locking. The in-loop frequency stability of repetition rate is about $1.6 \times 10^{-13}$ at 1 second integration time, limited by the measurement system; and it is inversely proportional to integration time in short term. Furthermore, using a common path $f$-$2f$ interferometer, the carrier envelope offset frequency of the frequency comb is obtained with a signal-to-noise ratio of 40 dB in 3 MHz resolution bandwidth. Stabilized carrier envelope offset frequency exhibits a deviation of 0.6 mHz at 1 second integration time.
**Keywords:** Optical frequency comb, Fiber laser, Frequency stabilization, Frequency instability
**PACS:** 42.55.Wd, 42.62.Eh, 42.65.Re

## 1. Introduction

Optical frequency combs, based on Ti:sapphire femtosecond laser, were invented at the end of 1990s [1,2]. Two scientists from Germany and USA won Nobel Prize for physics in 2005, due to this invention. Nowadays, Optical frequency combs are widely used in many applications, including attosecond science [3], spectroscopy [4], astronomical spectrograph [5], low-noise microwave generation [6], distance measurement [7], time and frequency transfer [8], optical clock development and application [9], and so on. Requirement of various applications drives scientists and engineers to develop new optical combs based on different laser sources, such as Cr:LiSAF [10], Er:fiber [11], Cr:forsterite [12], Er:Yb:glass [13], Yb:fiber [14]. Especially, Erbium doped fiber (Er:fiber) frequency combs attract much attention because of their compactness, reliability, low power consumption and directly covering telecommunication wavelength. As a result, Er:fiber frequency combs rapidly evolve, and their performance is as good as that of the best comb; the frequency stability of optical spectrum transfer can reach $\sim 1\times 10^{-17}$ at 1 s averaging time [15].

*Project supported by the National Natural Science Foundation of China (Grant No. 91336101 and 61127901), and West Light Foundation of The Chinese Academy of Sciences (Grant No. 2013ZD02).
†Corresponding author. E-mail:haifeng.jiang@ntsc.ac.cn

Since optical frequency combs were invented, ultra-high resolution optical frequency measurement has always been one of the most important applications. In order to meet the need of this application, some critical techniques were proposed to improve the frequency stability of the combs [16-19]. Optical frequencies of a comb can be expressed as $v_n = f_{ceo} + n f_r$ [20], where $f_{ceo}$ is the carrier envelope offset frequency, $f_r$ is the repetition rate, n is a integer of $10^5 \sim 10^6$. A small frequency fluctuation of $f_r$ can led to a significe frequency various of $v_n$. Consequently, $f_r$ definitely need controlling in a high bandwidth, to overcome the unexpected changes introduced by environmental noises or others. Traditionally, $f_r$ is controlled only with a piezo-transducer (PZT), whose response bandwidth is usually well below 10 kHz. To enlarge the control bandwidth of $f_r$, scientists in JILA (USA) firstly realized fast servo of the repetition rate (~230 kHz) by using an intra-cavity electro-optic modulator (EOM) [16]. In China, there are a few capable research groups, making efforts to develop fiber-based comb system with an intra-cavity EOM [21-23].

In this paper, we present the progress of an erbium-fiber-laser-based optical frequency comb for strontium optical clock frequency measurement. Its repetition rate ($f_r$) is about 209 MHz. In order to control $f_r$, an intra-cavity EOM and a long-range PZT are used as the feedback terminals simultaneously. Based on this configuration, $f_r$ is able to be stabilized continuously for more than one week. The response bandwidth of $f_r$ control is improved to megahertz, and the tunable range is 3 kHz. Moreover, the carrier envelope offset frequency ($f_{ceo}$) of the Er:fiber laser, with a signal-to-noise ratio (SNR) of 40 dB in 3 MHz resolution bandwidth (RBW), is detected by a common path $f$-$2f$ interferometer. After stabilization, $f_{ceo}$ exhibits a deviation of 0.6 mHz at 1 second integration time.

## 2. Experimental setup

Figure 1 shows the setup of the optical frequency comb, which is based on an Er:fiber femtosecond laser. The laser output is divided into three parts. One part is the optical frequency comb output for the subsequent application; the other two are used for detection and stabilization of $f_{ceo}$ and $f_r$ respectively.

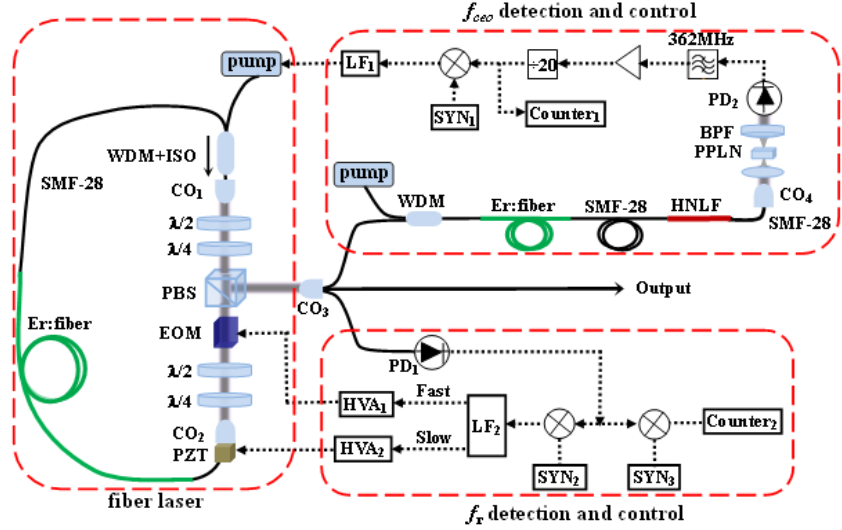

Fig. 1. Schematic of the Er:fiber femtosecond laser frequency comb. The thick solid lines represent free-space paths, the thin solid lines represent fiber paths and the dotted lines represent electrical paths. λ/2: Half-wave plate; λ/4: Quarter-wave plate; WDM: Wavelength division multiplexer; ISO: Isolator; PBS: Polarization beam splitter; CO: Collimator; PD: Photodetector; BPF: Bandpass filter; PPLN: Periodically poled lithium niobate; HNLF: Highly nonlinear fiber; LF: Loop filter; HVA: High voltage amplifier; SYN: Synthesizer.

**2.1 Er:fiber-based mode-locked laser**

As shown in Fig. 1, the Er:fiber femtosecond laser source has a ring cavity including four wave plates, a PBS, two collimators, a WDM with isolator, an EOM and fibers. A 2-cm PZT mounted on a stage is employed to control the cavity length. The cavity contains 39 cm SMF-28 fiber, 8 cm HI-1060 fiber and 39 cm highly erbium-doped fiber (Liekki ER110-4/125). Net dispersion of the cavity is supposed to be slightly negative, considering all components and fibers in the cavity. The laser is pumped with two single-mode 980 nm pigtailed diode lasers, whose power is up to 1 watt. By setting wave plates in proper angles, the laser can start mode-locking automatically while the pumping power is 1 watt. Unfortunately, the mode-locked laser works in a multi-pulses state in this condition. In order to avoid this situation, we decrease the pump power to around 350 mW. The output power is ~ 50 mW, and the pulse duration emitted from PBS is ~ 83 fs with a 3-dB optical spectrum width of ~ 50 nm.

Two feedback terminals, an EOM and a PZT, are employed to control $f_r$. The EOM is an 8-mm long lithium niobate crystal with 800 $fs^2$ dispersion at 1550 nm. To measure the response bandwidth of the EOM, we setup a Mach-Zehnder interferometer which contains the EOM in one arm. Then, we drive the EOM with an external radio-frequency (RF) source and monitor the phase response of the EOM at the output of the interferometer using a 2 GHz bandwidth photodetector (PD). The 3-dB bandwidth in magnitude is about 10 MHz, while the phase delay is close $\pi/2$. The change of $f_r$ caused by EOM is about 60 Hz monotonically while tuning offset voltage from -200 V to 200 V. And the tuning range with the PZT (from 0 V to 200 V) is measured to be around 3 kHz with a response bandwidth of hundreds Hz.

## 2.2 $f_r$ detection and stabilization

To detect the repetition rate, we sent a fraction of the output from the mode-locked laser into an InGaAs PD (EOT-3000A) with a bandwidth of 2 GHz. At the output of the PD, we obtain a current pulses train exhibiting $m \times f_r$, where $m$ is an integer, in frequency domain. Usually, we do not stabilize $f_r$ directly, but stabilize its high order harmonics, so as to reduce technical noise induced by the phase-locking process. However, no high frequency microwave synthesizer is available for the moment. Eventually, we lock the phase of the fourth harmonics of the repetition rate (838 MHz) to a reference signal.

The repetition rate is stabilized with a standard phase-locked loop technique. The phase detector is a RF mixer which produces an error signal indicating phase difference between $4f_r$ signal and the reference signal. Following the mixer, a home-made loop filter converts the error signal into negative feedback signal for driving EOM with high bandwidth and PZT with long range. Although the EOM has a response bandwidth well about 10 MHz, the control bandwidth of EOM control-loop is estimated to be 100 kHz - 200 kHz according to the response measurement results of the home-made loop-filter and high voltage amplifier. PZT control-loop has a bandwidth of below 100 Hz for compensating only temperature drift effect.

In addition, $f_r$ keeps being phase-locked while the box of the laser being hit with fist, because EOM control-loop cancels vibration influence efficiently. The peak-to-peak temperature fluctuation is about 1 °C in the laboratory, and well-below 1 °C in the laser box, so that PZT control-loop can eliminate the thermal drift effect independently. Extra thermal control is not required. As a result, $f_r$ of the laser can be continuously phase-locked onto the reference signal over a week.

## 2.3 $f_{ceo}$ detection and stabilization

We use *f-2f* interference method to produce $f_{ceo}$ signal; and lock $f_{ceo}$ to an extra-reference (radio-frequency RF) signal in phase by controlling ring laser's pumping power.

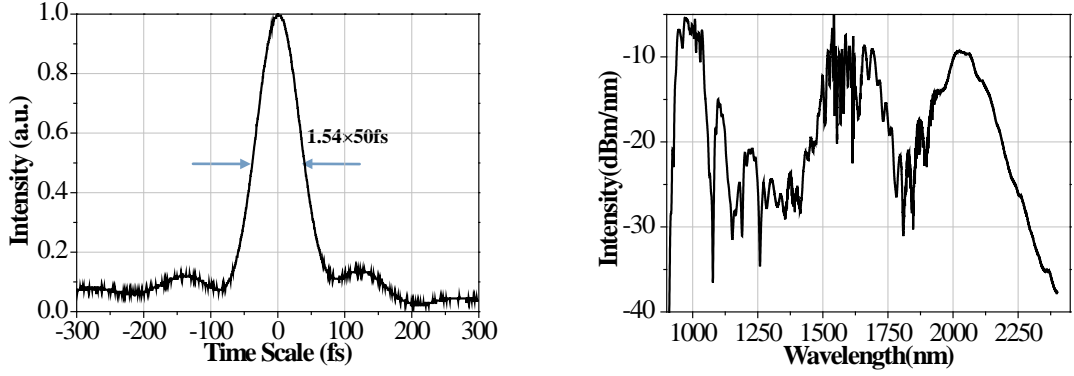

Fig. 2. (a) The measured intensity autocorrelation of the compressed pulse (b) The octave-spanning supercontinuum produced with HNLF.

In order to perform *f-2f* interference, we first pre-chirp seed pulses for reducing the pulse peak power, aiming to decrease nonlinear effect in amplification stage. Secondly, the pre-chirped pulses are delivered into an erbium-doped fiber amplifier with a selective optimum gain of ~ 22 dB. Thirdly, compress laser pulse width by de-chirping the amplified pulse with SMF-28 fiber, yielding ~ 50 fs pulse duration as shown in Fig. 2(a). Fourthly, feed these narrow pulses into a short highly nonlinear fiber (HNLF), resulting in an octave spanning supercontinuum (SC) as shown in Fig. 2(b). This figure gives the optical spectrum of the SC spanning from 950 nm to 2250 nm for 25 dB bandwidth. Most of the energy shifts into two solitons with 1 μm and 2 μm wavelength respectively. It is worth noting that these two solitons are exactly in the right wavelength for *f-2f* interference. After that, direct all optical power through a 1-mm thick PPLN crystal for frequency-doubling. Finally, we pick-up 1μm optical power with a 10 nm bandpass filter and obtain beatnote signal of *f-2f* with an InGaAs PD.

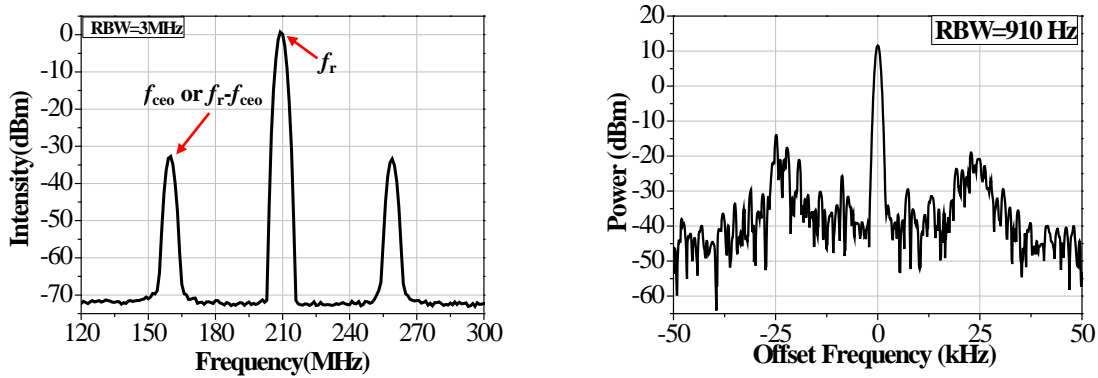

Fig. 3. (a) Spectrum of the *f-2f* interference in 3 MHz RBW (b) Spectrum of the stabilized $f_{ceo}$ signal in 910 Hz RBW.

Figure 3(a) shows the spectrum of the *f-2f* interference with 3 MHz RBW, demonstrating a

SNR of $f_{ceo}$ to be 40 dB. The linewidth of the free-running $f_{ceo}$ beat is about 500 kHz observed in fine RBW. The SNR is about 20 dB higher than the previously reported Er:fiber domestic system [22]. High SNR enables the system to generate a clean feedback signal for controlling the pumping power and stabilizing $f_{ceo}$ on a reference frequency provided with a synthesizer. A 20-times frequency divider is used to improve the phase-locking tolerance, making the phase-locked loop more robust. The control bandwidth of the loop is about 25 kHz indicated by the gain-bump as shown in fig. 3(b). The accumulated noise power, from 50 kHz to 10 Hz away from the carrier, is about 1% indicating that the linewidth of stabilized $f_{ceo}$ should be well below 1 Hz. More accurate evaluation will be shown later with frequency stability.

The longest continuously phase-locking time of $f_{ceo}$ is more than a day. However, some unknown noises coupled from electric ground (introduced by other events) disturb the phase locked loop from time to time. We suppose that $f_{ceo}$ can be stabilized much longer if the electric ground gets clean.

## 3. Results and discussion

Frequency stability of the optical frequency combs is one of the most important parameters. Measurement of comb's absolute (out-of-loop) frequency instability requires two independent systems which are not available now. Here, we measure only in-loop frequency instability which does not include parts of fiber noise in comparison with the out-of-loop one.

3.1 in-loop frequency stability of $f_r$

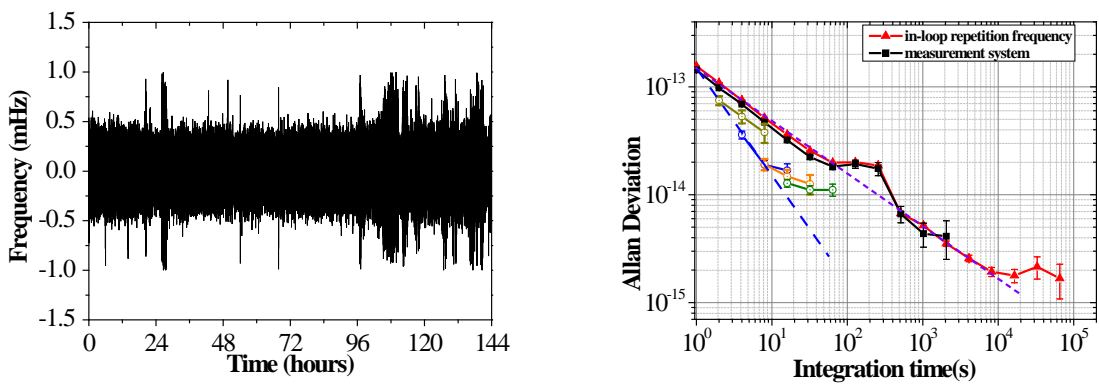

Fig. 4. (a) Residual fluctuations of the 4th harmonic of the repetition rate; (b) In-loop frequency stability of $4f_r$, measured with different gate (solid red triangle for 1s, empty round for 2s, 4s, 8 s and 16 s) time and measurement noise floor (solid black square).

As shown in Fig. 1, down-converted frequency of $4f_r$ at 1 MHz is recorded with a frequency counter (Agilent 53230A). Fig. 4(a) shows the frequency fluctuation of $4f_r$ measured over a week.

Note that the gate time is 1 second and the dead-time is about 0.2 second. The result exhibits a root mean square (RMS) value of about 0.1 mHz. The frequency stability ($\Delta f_r/f_r$) shown in Fig. 4(b) is about $1.6\times10^{-13}$ at 1 second integration time. However, this result is limited by the noise floor, as shown with solid black squares, which is measured by using the same counter recording a 1 MHz reference signal. On the other hand, the frequency stability is inversely proportional to $\tau^{1/2}$, where $\tau$ is the integration time, because the dead-time decouples the data relationship in measurement system and converts the white phase noise into white frequency noise [24]. For a phase locked loop, the stabilized signal should exhibit white phase noise, and its frequency stability should be inversely proportional to $\tau$. This is verified by eliminating dead-time effect with different gate times as shown with a blue dashed line in Fig. 4(b). Note that the first point of calculated Allan Deviations has no dead-time effect. Therefore, the frequency stability of $f_r$ is about $1.6\times10^{-13}$ at 1 second integration time and rolls down with a slope of $1/\tau$ for short term stability.

### 3.2 in-loop frequency stability of $f_{ceo}$

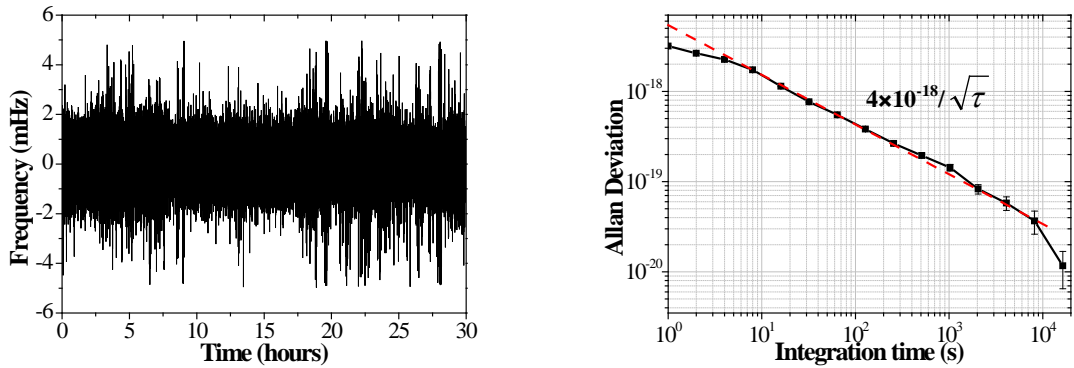

Fig. 5. (a) Residual fluctuation of the stabilized in-loop CEO frequency; (b) Optical comb frequency instability induced by in-loop $f_{ceo}$ fluctuation.

The RMS of $f_{ceo}$ fluctuation is about 0.6 mHz at 1 second integration time, and the peak-to-peak fluctuation is about 10 mHz as shown in Fig. 5(a). Considering the optical frequency is about 200 THz, the corresponding optical comb frequency stability induced by in-loop $f_{ceo}$ fluctuation is about $4\times10^{-18}/\tau^{1/2}$ as shown in Fig. 5(b).

It is clear that the frequency stability induced by in-loop $f_{ceo}$ fluctuation is negligible, even compared with the highest stable atomic clock with a stability of $3.2\times10^{-16}/\tau^{1/2}$ [25, 26].

### 5. Conclusion and perspective

In summary, we demonstrate a home-made Er:fiber based optical frequency comb with an intra-cavity EOM. To the best of our knowledge, this optical frequency comb is the first domestically-made one with the fast $f_r$ controller in the laser cavity. Both $f_r$ and $f_{ceo}$ of the comb can be easily stabilized continuously over a long time, thanks to the high bandwidth frequency control and high SNR of frequency signals detection. The frequency instability of the system is well below $1.6\times10^{-13}$ @ 1s and rolls down with a slope of $1/\tau$ for short term. In the future, we will continuously optimize the system, and expend its operation frequency to strontium clock transition (698 nm) for frequency measurement.